%% file: 0_Main.tex
\documentclass[aip,twocolumn, jmp, amsmath,amssymb, reprint,floatfix,]{revtex4-1}

\usepackage{import}
\usepackage{graphicx}
\usepackage{dcolumn}
\usepackage{bm}
\usepackage{braket}
\usepackage{enumitem}
\usepackage{float}
\usepackage[section]{placeins} 
\usepackage{booktabs}

\begin{document}
\title{Effects of perturbation order and basis set on alchemical predictions}
\author{Giorgio Domenichini, Guido Falk von Rudorff and O.Anatole von Lilienfeld}
\email{anatole.vonlilienfeld@unibas.ch}
\affiliation{Institute of Physical Chemistry and National Center for Computational Design and Discovery of Novel Materials (MARVEL), Department of Chemistry, University of Basel, 4056 Basel, Switzerland}

\begin{abstract}
Alchemical perturbation density functional theory 
has been shown to be an efficient and computationally inexpensive way to explore chemical compound space. 
We investigate approximations made, in terms of atomic basis sets and perturbation order,
introduce an electron-density based estimate of errors of the alchemical prediction, 
and propose a correction for effects due to basis-set incompleteness. 
Our numerical analysis of potential energy estimates, and resulting binding curves, is based on CCSD reference results, and is limited to all neutral diatomics with 14 electrons (AlH ... N$_2$). 
The method predicts binding energy, equilibrium distance, and vibrational frequencies of neighbouring out-of-sample diatomics with near CCSD quality using perturbations up to 5$^{th}$ order.  
We also discuss simultaneous alchemical mutations at multiple sites in benzene. 
\end{abstract}
\maketitle

\import{./}{1_Introduction.tex}
\section{Methods}
\import{./}{2A_CompProcedures.tex}
\import{./}{2B_Sources_prediction_errors.tex}

\import{./}{2C_define_bserror}
\section{results}
\import{./}{3_0_this_work_details.tex}
\import{./}{3A_scatter.tex}
\import{./}{3B_ao_error.tex}
\import{./}{3C_ao_correction.tex}
\import{./}{3D_Truncation_error.tex}
\import{./}{3E_Vibrational.tex}
\import{./}{3F_benzene_pyridine.tex}
\import{./}{4_conclusion.tex}

\section*{Acknowledgement}
O.A.v.L. acknowledges funding from the Swiss National Science foundation (200021\_175747) and from the European Research Council (ERC-CoG grant QML). This work was partly supported by the NCCR MARVEL, 
funded by the Swiss National Science Foundation.

\section{References}
\bibliography{bib_files/BasisRef,bib_files/Alchemy_vLg,bib_files/Alchemy_Others,bib_files/software,bib_files/Other_works,bib_files/QML}
\medskip
\FloatBarrier

\end{document}

%% file: 1_Introduction.tex
 \section{Introduction}
Chemical space, the ensemble of all possible molecule that constitute matter, is unfathomably large. The number of molecules in chemical space is estimated to be higher than $10^{60}$ just considering small organic molecules \cite{ccs,ccs2}.
The size of this problem makes it impossible to enumerate and assess a significant portion of chemical space with standard quantum chemistry methods alone. Therefore, modern approaches in materials design have the need for computationally less demanding, yet sufficiently accurate methods \cite{ccsLilienfeld}. 
Promising results were obtained for instance through quantum machine learning models \cite{qml1,qml2,qml3,qml4,qml5,qml6}. Provided that the machine has been trained on a sufficiently large number of molecules, quantum machine models can provide fast yet accurate predictions. For example, in 2017 machine learning models were trained to predict multiple electronic properties for thousands of organic molecules reaching DFT accuracy in milli-second prediction time\cite{Faber1,Faber2}.

Alchemical perturbation density functional theory (APDFT) aims at a similar goal, namely to obtain a rapid screening of chemical space. The approach is fundamentally different though: While machine learning models require thousands or millions of training instances in order to interpolate towards similar systems, APDFT relies on a single explicit calculation on a reference system to give approximate, but still accurate predictions for properties of a multitude of isoelectronic and similar target compounds. \cite{Lilienfeld2009,Chang2018,benzene2013,Lilienfelds_bn_doped_graphene,c60_2018}
APDFT yields consistent predictions of both energies and electron densities\cite{apdft}. Its reliability has been shown in several applications such as energies in BN-doped aromatic systems and non-covalent interactions thereof\cite{apdft}, decomposition of energy contributions\cite{Rudorff2019a} or estimation of deprotonation energies\cite{Rudorff2020,Cardenas}. 
Through the use of pseudopotentials, quantum alchemy can be a powerful tool in solid state chemistry \cite{marzari_94,Lilienfeld2009,Lilienfeld2010,Solovyeva_crystals_2016,Chang2018}, with possible application in catalysts design \cite{Keith2017,Keith2019,Keith2020} . 
In recent applications, it has been shown that APDFT---if applied to high level CCSD \cite {ccsd1,ccsd2,ccsd3} calculations for the reference molecule---can outperform\cite{apdft} widely used methods in computational chemistry (e.g.  HF\cite{1935,slater1951}, MP2\cite{Moller1934} and DFT\cite{becke1993,lee1988,vosko1980,pbe}) 
This approach holds the promise to shift the computational cost from many medium-quality calculations throughout chemical compound space to a few select high-quality calculations which serve as a subsequent basis to obtain alchemical estimates of electronic observables for a combinatorially larger number of molecules in one shot. 

APDFT is straight-forward: The non-degenerate electronic ground state in the Born Oppenheimer approximation $E_0(\mathbf{R}_I,Z_I,N_e)$ is a continuous function of (i) the positions of the nuclei $\mathbf{R}_I$, (ii) the nuclear charges $Z_I \le 0$, and (iii) the number of electrons $N_e$ of a molecule.
Partial derivatives  of observables such as $E_0$ with respect to the nuclear charges (constant $\mathbf{R}_I$ and $N_e$) are called alchemical derivatives and can be used to perturb a reference molecule and obtain efficient estimates of solutions for multiple target molecules via the Taylor series expansion. 
As such, iso-electronic alchemical changes take place within the same Hilbert space populated by all the solutions to Schr\"odinger's equation for {\em any} combination of real values for $\{Z_I,{\mathbf{R}_I}\}$.  Obviously, comparison to experimental realizations are only meaningful when all nuclear charges assume integer positive values. 

Due to the Hellmann-Feynman theorem, the first alchemical derivative can be obtained analytically and typically at negligible additional cost. \cite{4DEDF,Levy1978} 
The second derivatives of the ground state energy with respect to $N_e$, $R$ and $Z$ and their mixed partial derivatives constitute a unified Hessian matrix \cite{anm} which is a generalization of the "geometrical" Hessian matrix, in which only the derivatives with respect to the nuclear coordinates are included, and which can be diagonalized in order to generate a complete basis in which chemical compound space can be expanded.  
Rigorously rooted in Rayleigh-Schr\"odinger-perturbation theory, APDFT is an approximate method in practice. 
While other (uncontrolled) approximations are common in DFT, the approximate nature of APDFT has not yet been explored in full. As such, it is highly desirable to obtain at least an estimate of the error associated with alchemical estimates for given reference calculations and given perturbation orders. 
Control of prediction errors represent an important goal for computational materials design efforts, 
as meaningful trade-offs between a calculation's cost and its accuracy can be established. 

Note that while in principle the concept of alchemical prediction applicable to all electronic properties\cite{apdft}, here we focus on the ground state energy $E_0$. 
More specifically, we give a comprehensive discussion of the sources of errors from a practical point of view, i.e. also including those sources of errors which are not necessarily inherent to APDFT \textit{per se} but rather result from restrictions in present quantum chemistry codes. We explain the relevance of the various sources of errors and give a correction for the error in energy resulting from atomic basis sets (validated on pyridine, bipyridine and triazine). Based on a set of diatomics, we calibrate a measure for the error of quantum alchemy. Finally, we demonstrate that APDFT calculations of vibrational frequencies in dimers can be nearly as good as the reference method employed (CCSD), and more accurate than common DFT approximations. 

%% file: 2A_CompProcedures.tex
\subsection{\label{sec:level1}Alchemical Perturbation Density Functional Theory}
Following earlier work\cite{apdft} where details on the derivation are given, we  define the alchemical coordinate $0 \le \lambda \le 1$ as a linear transformation of the nuclear charges from the reference vector of nuclear charges $Z^\text{R}$ to the target $Z^\text{T}$. For atom $I$:
\begin{equation}
  \begin{aligned}
  Z_I(\lambda) &\equiv Z_I^\text{R} + \lambda (Z_I^\text{T}-Z_I^\text{R}) = Z_I^\text{R} + \lambda  \Delta Z_I
\end{aligned}
\end{equation}

For isoelectronic changes, the total electronic ground state energy of the molecule is continuous and differentiable in $\lambda$. \cite{Lilienfeld2009} The function $E(\lambda)$ can be expanded as Taylor series centered at $\lambda=0$, i.e. at the reference molecule. With the energy of the reference molecule $E^\text{R} = E(\lambda=0)$ and its derivatives, we can approximate the energy of the target molecule $E^\text{T}= E(\lambda=1)$.

\begin{equation}
  \begin{aligned}
   E^\text{T} &= \sum_{n=0}^{\infty}{ \frac{1}{n!} \left.\frac{\partial^n E(\lambda)}{\partial \lambda^n}\right|_{\lambda=0}}= E^\text{R} + \sum_{n=1}^{\infty}{ \frac{1}{n!} \left.\frac{\partial^n E(\lambda)}{\partial \lambda^n}\right|_{\lambda=0}}
\end{aligned}
\end{equation}
The first derivative can be evaluated via the Hellmann-Feynman theorem\cite{HF_theorem_Feynman}: 
\begin{equation}
\begin{aligned}
\label{firstderiv}
\left.\frac{\partial E}{\partial \lambda}\right|_{\lambda=0} & = \bra{\psi_\text{R}}\hat{H}_\text{T} - \hat{H}_\text{R} \ket{\psi_\text{R}} \\
\end{aligned}
\end{equation}
which can be evaluated analytically for any wavefunction-based or density-based quantum chemistry method. Given the electron density, one finds~\cite{Lilienfeld2009}:
\begin{equation}
\begin{aligned}
\label{eq_1st_deriv_int}
\frac{\partial E}{\partial \lambda} & = \int_\Omega d \mathbf{r}(v_\text{T}(\mathbf{r})-v_\text{R}(\mathbf{r})) \rho(\mathbf{r},\lambda)
\end{aligned}
\end{equation}
with the external potentials $v_\text{R}$ and $v_\text{T}$ corresponding to reference and target systems, respectively.
From this it is clear that Eq.~\ref{eq_1st_deriv_int} can be written in terms of higher order perturbations of the electron density alone~\cite{apdft}
\begin{equation}
\begin{aligned}
\label{rho_deriv_eq}\frac{\partial^{n+1} E}{\partial \lambda^{n+1}}
=\int_\Omega d \mathbf{r}(v_\text{T}(\mathbf{r})-v_\text{R}(\mathbf{r})) \frac{\partial^n \rho(\mathbf{r},\lambda)}{\partial \lambda^n} 
\end{aligned}
\end{equation}
Combining Eqs.~\ref{rho_deriv_eq} and 2, one obtains
\begin{equation}
E^\text{T} = E^\text{R} + \int_\Omega d \mathbf{r}(v_\text{T}(\mathbf{r})-v_\text{R}(\mathbf{r}))
\tilde{\rho(\mathbf{r})} 
\end{equation}
---an orbital free density functional which is exact provided that (i) $E^\text{R}$ is exact and that (ii) the Taylor expansion converges, and where the averaged density $\tilde{\rho}$ is given by
 $\tilde{\rho(\mathbf{r})} = \sum_{n=1}^\infty \frac{1}{n!} \frac{\partial^{n-1} \rho(\mathbf{r},\lambda)}{\partial \lambda^{n-1}}|_{\lambda = 0} $.

The integrals \ref{eq_1st_deriv_int} and  \ref{rho_deriv_eq} can be obtained from numerical integration by projecting the one particle electron density on an integration grid
or analytically by contracting the one particle density matrix with the nuclear attraction operator in atomic orbital basis. Both method are equivalent provided a suitable integration grid is used.

In this work, we evaluate the derivatives of the electron density via finite differences on an evenly spaced five point stencil with $\Delta\lambda= 0.05$. Based on previous results, this stencil yields good numerical stability\cite{apdft}. The stencil coefficients have been calculated\cite{FDcoeffs} such that the leading error term is $\mathcal{O}(\Delta\lambda^5)$.
In practice, the infinite sum from the Taylor expansion is truncated after the perturbation order $n$. Estimates made using all terms up to order $n$ are denoted by APDFT$n$.

%% file: 2B_Sources_prediction_errors.tex
\subsection{Sources of prediction errors} 
Calculating the alchemical derivatives via finite differences requires the evaluation of the density derivatives in the basis set of the reference. While plane wave calculations use the same basis set for target and reference, atom-centered orbital basis sets do not. Atomic orbitals are commonly chosen because are a reliable and inherently local way to represent the wave function in molecules.

Consequently, using atom-centered basis sets means that APDFT derivatives actually build up the energy of the target molecule in the basis set of the reference molecule. With atomic orbital basis sets being optimized for each element, this results in a potentially noticeable error in the target energy.

The energy of the target molecule with the basis set of the reference $E^\text{T[R]}$ is in general different and usually higher than the energy of the target with its  optimized basis set $E^\text{T[T]}$. We define the energy error due to the use of the reference basis set $\Delta E_\text{BS}$ as the difference between these energies.
\begin{equation} \label{dEbs_def}
     \Delta E_{\textrm{BS}} \equiv E^\text{T[R]} - E^\text{T[T]}
\end{equation}

In contrast to this error, the truncation of the Taylor series to a certain number of leading terms introduces other errors. This error contribution is equal to the difference between the predicted energy $E^\text{APDFT}n$ and the energy of the target molecule $E^\text{T[R]}$ evaluated with the basis set of the reference molecule:
\begin{equation}
     \Delta E_\text{trunc} \equiv E^\text{APDFT} -E^\text{T[R]}
\end{equation}

Besides these higher order terms, practical implementation details commonly introduce two more sources of error: i) the error in the estimation of derivatives due to the finite difference scheme and ii) the error from numerical integration of the Coulomb interaction with the perturbed densities. These two errors can potentially be minimized by choosing a more numerically stable integration grid or a more accurate finite difference stencil.

As such, we view the intrinsic error of APDFT$n$ as the sum of two dominating contributions, one due to the truncation of the Taylor series at order $n$, and the other one due to using the basis set of the reference molecule:
\begin{equation}
    \begin{aligned}
     \Delta E_{\textrm{APDFT}}&=E^{\textrm{APDFT}}-E^{\textrm{T[T]}}\\
     &=   E^{\textrm{APDFT}}-E^{\textrm{T[R]}}+E^{\textrm{T[R]}}-E^{\textrm{T[T]}}\\
     &= \Delta E_{\textrm{trunc}}+\Delta E_{\textrm{BS}}
    \end{aligned}
\end{equation}
While the truncation order can be controlled through increase of $n$ (as long as the Taylor expansion converges), basis set effects have been studied less.

%% file: 2C_define_bserror.tex
\subsection{Correction for Atom-Centered Orbital Basis Sets} \label{bsc_def_sect}
For small to medium atomic basis sets, the error contributions $\Delta E_\text{BS}$ which arise from representing the density derivatives in the basis set of the reference molecule dominate the total deviation from a self-consistent evaluation of the target molecule. Technically, the exact error $\Delta E_\text{BS}$ can be recovered by including alchemical derivatives w.r.t. the basis set coefficients\cite{zachara2012}. This, however, would substantially increase the computational cost as we go to higher orders in the perturbation. While the error vanishes in the complete basis set limit (see numerical evidence below), a correction constitutes a worthwhile alternative, enabling accurate absolute energy estimates while remaining cost-effective. Note that relative energies, such as atomization energies, are less affected by this problem due to a cancellation of all those error-contributions which are independent of the atomic environment. 

To motivate the correction, we observe that the largest contribution to the total electrostatic energy of the molecule is given by the core orbitals, thus we can also assume that the main contribution in $\Delta E_\text{BS} $ is due to the change in the core orbitals' coefficients. This is both because the core orbitals have a larger contribution to the total energy than valence orbitals and because they are less flexible than their valence counterparts.\\

If $\Delta E_\text{BS} $ is predominantly unaffected by the chemical environment we can approximately estimate it using the full dissociation limit, i.e.~the individual free atoms. For each  atom $I$, an atomic contribution $\Delta E_\text{BS,I}$ is collected, the sum of which amounts to our approximate correction of $\Delta E_\text{BS}$

\begin{equation}
\Delta E_\text{BS} \approx \sum_{I} E^\text{T[R]}_I-E^\text{T[T]}_I \equiv  \sum_{I} \Delta E_{\text{BS},I}
\end{equation}
For every atom, $E_{\text{BS},I}$ is the difference between the free atom energy with the basis set of the reference element for that site minus the energy of the free atom with its proper basis set.\\

This correction has no effect on the shape of the potential energy surface, it does improve total energies with limited and constant additional computational effort, namely a single free atom calculation for each target element is required. For example, in order to study BN-doping in $sp^2$-hybridized carbon sites, four such elemental calculations are necessary, regardless of the number of target compounds: $E^\text{B[B]},E^\text{B[C]},E^\text{N[N]},E^\text{N[C]} $. Here, all free atom calculations have been performed in their their lowest electronic spin state. While certainly of interest, the study of alchemical changes involving other spin-state combinations is so substantial that it warrants a separate publication.

%% file: 3_0_this_work_details.tex
As test case for this work, we consider all pairwise APDFT predictions that can be made within the  series of neutral diatomic molecules with 14 electrons (HAl, HeMg, LiNa, BeNe, BF, CO, and NN).  This set not only covers the elements most relevant to organic chemistry, but also allows to assess APDFT predictions between many different elements. All of these molecules are isoelectronic, therefore the energy of any molecule in this set can be estimated via alchemical transformation from any other molecule in this set. To account for contributions from non-equilibrium geometries, we consider 20 bond lengths from 1.3 to 3.2\,Bohr.

Since the accuracy of APDFT depends significantly on the basis set, we performed all calculations for major basis set families. Within each family, we chose two representatives: one smaller and another more expanded basis set. This allows to compare the performance of APDFT both within and between basis set families. The employed basis sets are STO-3G, STO-6G\cite{stos}, Pople's split valence (\mbox{3-21G}, \mbox{6-31G*})\cite{321g,631g}, Dunning's  correlation consistent (cc-pVTZ, aug-cc-pVQZ)\cite{Dunning_1989,Dunning_1993} and the Karlsruhe (def2-TZVP, def2-QZVPP) \cite{def2tz,def2qz}  basis sets. 

The data set for the prediction up to APDFT5 within the 14 electrons diatomic series is made available under open access.\cite{domenichini_giorgio_2020_3959316}

All calculations employ CCSD in the frozen core approximation using a RHF reference. For comparison in section \ref{pes_sect}, DFT calculations with B3LYP and PBE functionals were performed. Unless otherwise noted, we used the quantum chemistry software MRCC\cite{mrcc,mrcc_article}. 

The calculations of the response density matrix in atomic basis for benzene in section \ref{rings_sect} were implemented using PySCF\cite{pyscf}.

%% file: 3A_scatter.tex
\subsection{Error From Reference Basis Set}
\begin{figure}[h!]
\includegraphics[width=\linewidth]{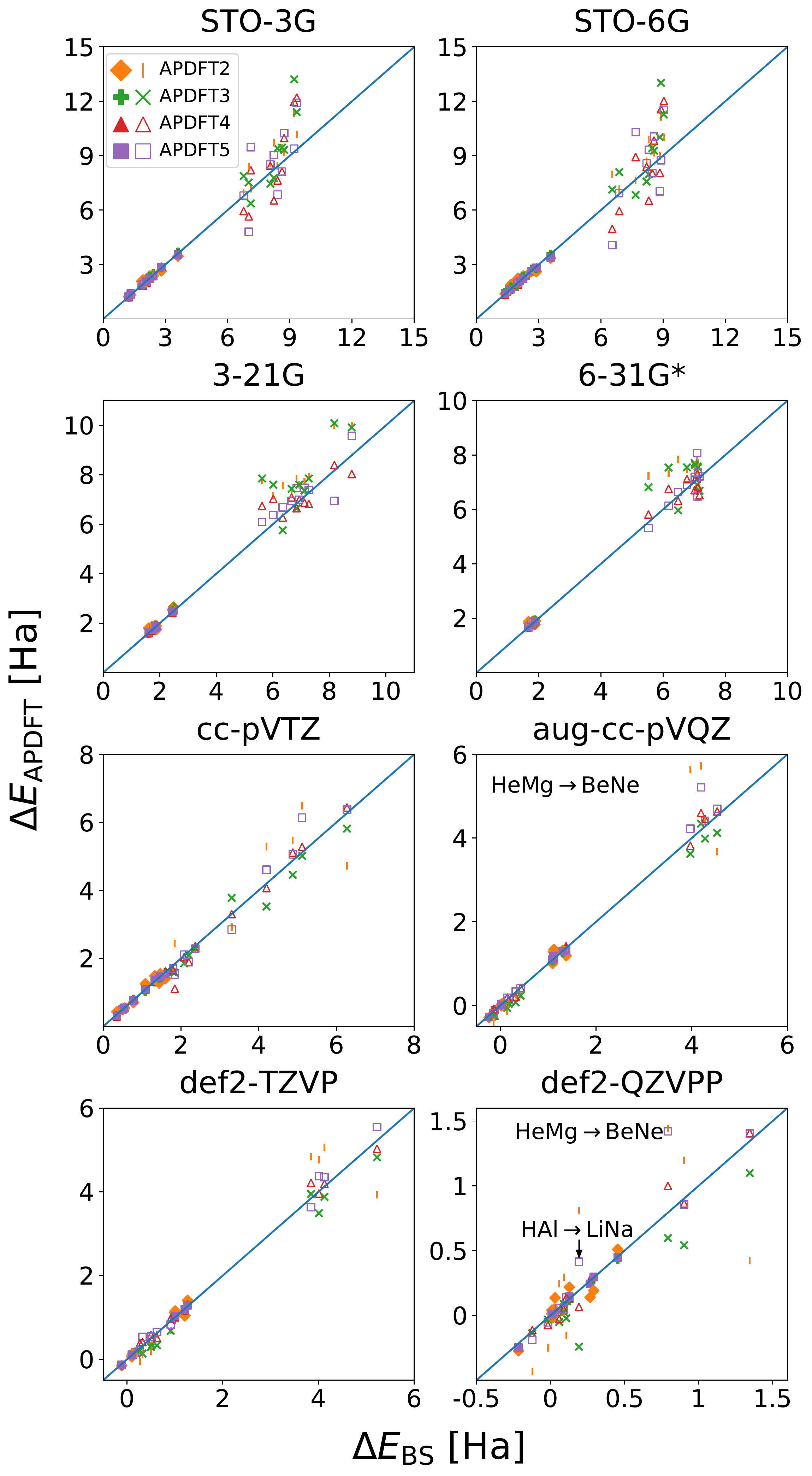}
\caption{The scatter plot shows the correlation between $\Delta E_\text{BS}$ and $\Delta E_{\text{APDFT}}$ for alchemical transmutations with $|\Delta Z|= 1$ (solid markers) and $|\Delta Z| = 2$ (hollow markers). Every pair of reference-target molecules within the 14 electron diatomic series is represented by a point for each APDFT order. The error shown is the median of all considered geometries. }
 \label{scatter_plt}
\end{figure}

For all alchemical absolute energy estimates within the diatomic iso-electronic series ($N_e$=14) for which the nuclear charges of reference and target molecule differ at most by 2, Figure~\ref{scatter_plt} shows the comparison between the basis set error $\Delta E_\text{BS}$ and the total error for all the different contributions for APDFT perturbation orders ranging from 2 to 5. Each panel corresponds to a different basis-set. 
For all basis sets and perturbation orders, 
we see substantially larger errors for $\Delta Z = \pm 2$ than for $\Delta Z = \pm 1$.
This is not surprising since the former also constitutes a much larger perturbation than the latter.
In view of the scales and the correlation between the overall APDFT error and the basis set contribution thereto, this suggests that the basis set contribution constitutes the dominating source of error. Moreover, the results indicate that the contribution is highly systematic especially when it comes to the smaller perturbation ($\Delta Z = \pm 1$). 

For some basis sets (3-21G, 6-31G*, aug-cc-pVQZ), we observe clusters of errors, due to $\Delta E_\text{BS}$ being mostly constant for neighboring elements. In the case of aug-cc-pVQZ, the cluster is subdivided, as the contribution for $\Delta E_\text{BS}$ for elements within the second period is smaller than the contribution for those in the third period.
Also as $|\Delta Z|$ increases, we see less correlation between the total error and the basis set contribution. This is predominantly caused by the truncation of the Taylor series as shown by the higher order predictions being much more consistently off. In rare cases those two contributions become comparable in magnitude.

The convergence with expansion orders is slower with the minimal STO or Pople basis sets. Interestingly, one consequence is that higher order energies are not always an improvement over lower orders. As shown below, we attribute this to the representability of the alchemical density derivatives in those basis sets. Consequently, the overall magnitude of the errors in these basis sets is significantly higher than for e.g. the Karlsruhe basis sets, and we note that due to such erratic behavior, the use of APDFT in small basis-set calculations is {\em not} to be recommended. 

In the cases of aug-cc-pVQZ, def2-TZVP and def2-QZVPP we obtained some negative basis set error. Those correspond to the alchemical transformations BeNe$\xrightarrow{}$ HeMg, BeNe$\xrightarrow{}$LiNa, BF$\xrightarrow{}$LiNa. In these transformations, the total number of core electrons decreases from target to reference. Since we used frozen core CCSD,a reduction of the correlation energy is observed, consequently leading to a total basis set error of negative sign.

For some noticeable outliers in Figure \ref{scatter_plt}, HeMg$\xrightarrow{}$BeNe and HAl$\xrightarrow{}$LiNa), even fourth order contributions for the largest basis set still yield significant residual differences. Note that in all these cases, an atom from the first period is used as a reference for an atom from the second period with a substantially increased electronic extend. 
Consequently, we must caution against the use of APDFT across periods when using nuclear charges only. 

%% file: 3B_ao_error.tex
\subsection{Quantifying the Basis Set Derivative Error} \label{AOerror}
\begin{figure}
\includegraphics[width=\linewidth]{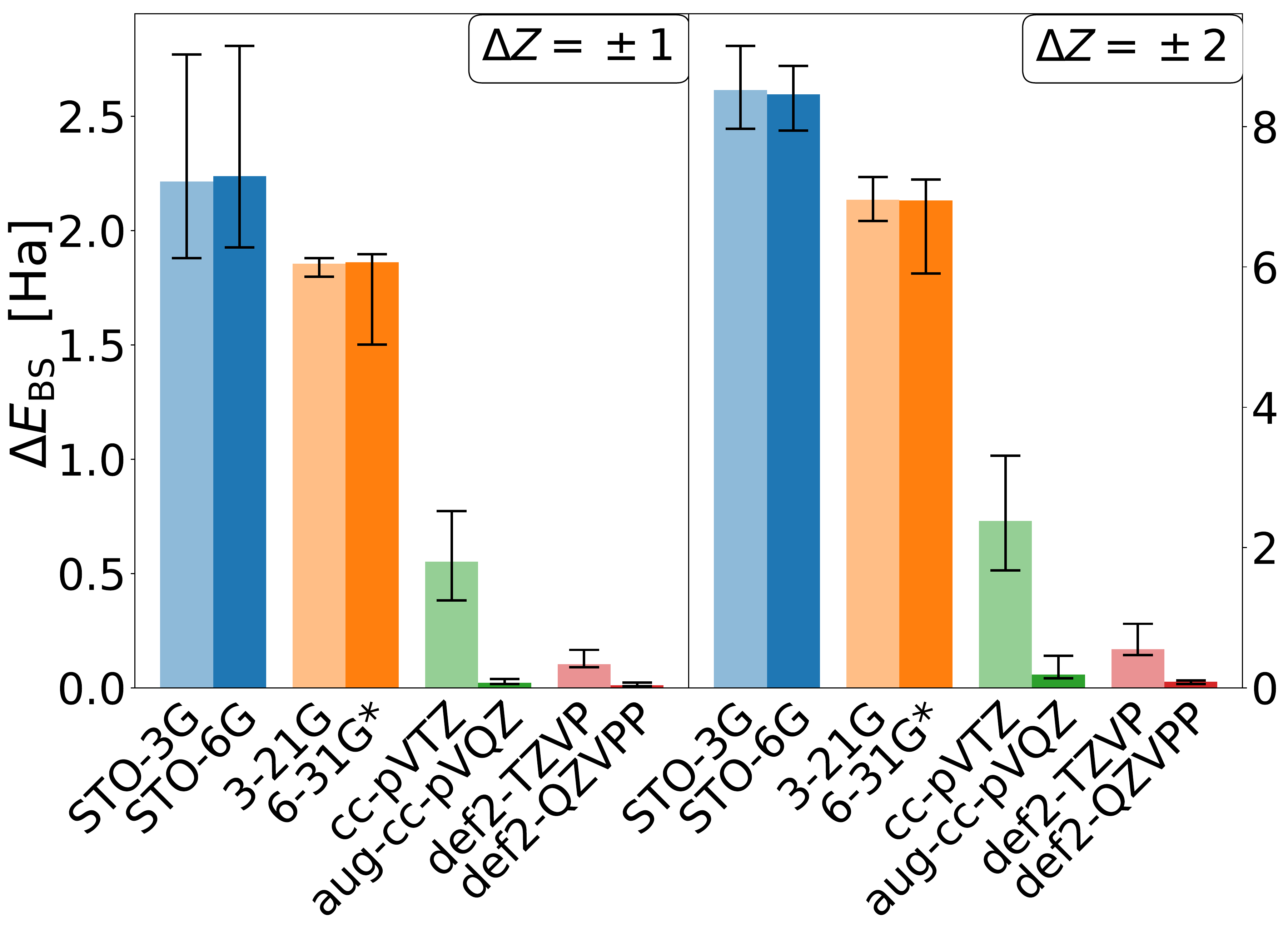}
\caption{Dependence of $\Delta E_{\text{BS}}$ on basis set type and size.
For alchemical transmutations within diatomics containing elements of the second period (BeNe, BF, CO, and NN) grouped by $|\Delta Z|$ are shown the median, the 10$^\text{th}$ and 90$^\text{th}$ percentile of $\Delta E_{\text{BS}}$. }
\label{bs_error_stat}
\end{figure}

\begin{table}
\begin{tabular}{lrrr}
\toprule   
Basis set &  \ \ \ \ $\Delta Z=\pm 1 $ & \ \ \ \ $\Delta Z =\pm 2$ & \ \ \ \  HAl $ \leftrightarrow $ HeMg \\
\hline
\midrule 
        STO-3G\rule{0pt}{12pt}  &       2215 &     8519 &    1684 \\
        STO-6G &       2238 &     8460 &    1687 \\
         3-21G &       1855 &     6956 &      1775 \\
        6-31G* &       1862 &     6947 &     1762 \\
      cc-pVTZ &        553 &     2382 &      1260 \\
  aug-cc-pVQZ &         23 &      190 &        1144 \\
    def2-TZVP &        105 &      552 &    1081 \\
   def2-QZVPP &         12 &       89 &     100 \\
\end{tabular}
\label{tab_bse}
\caption{Median of $\Delta E_{\text{BS}}$[mHa] for second period diatomics with $\Delta Z = \pm 1,2$ and for the pair HAl, HeMg ($\Delta Z = \pm 1 $).}
\end{table}

All sources of error depend on the basis set chosen for the reference calculation. By and large, the basis set error tends to be smaller for more expanded basis sets, approaching zero in the complete basis set limit. Since for some dimer pairs in this work the number of core electrons is not constant under alchemical transformation which can lead to an additional error for frozen-core CCSD, we  divided our dimer set into two subsets: HAl, and HeMg as well as BeNe, BF, CO, and NN. Within each subset, the number of core electrons remains constant under alchemical transformation between any set element. 
As shown in Figure~\ref{bs_error_stat} and Table \ref{tab_bse}, we find Karlsruhe def2 basis sets to be the most reliable in the context of alchemical changes as they outperform any other basis set family, in particular considering their number of basis functions. \\
The Karlsruhe basis set def2-TZVP has a basis set contraction scheme (11s,6p,2d,1f) $\rightarrow$ [5s,3p,2d,1f] similar to the cc-pVTZ basis set (10s,5p,2d,1f) $\rightarrow$ [4s,3p,2d,1f]. Even though they differ only by one $s$ base function the error for second row elements in cc-pVTZ is five fold higher than in def2-TZVP. \\
If we look at the quadruple-zeta basis set, def2-QZVPP with a second period contraction scheme of (15s,8p,3d,2f,1g) $\rightarrow$ [7s,4p,3d,2f,1g] is a substantially smaller basis set than aug-cc-pVQZ with a contraction scheme of (16s,10p,6d,4f,2g) $\rightarrow$ [9s,8p,6d,4f,2g]. Despite def2-QZVPP being smaller in number of basis functions, it performs better than aug-cc-pVQZ for APDFT, with typically half the error for $\Delta E_\text{BS}$.
The reasons for this may be found in the optimization procedures of these basis sets. For example, the first order energy derivative for vertical changes by virtue of the chain rule can be written as
\begin{align}
    \frac{\partial E}{\partial \lambda} = \sum_I \frac{\partial E}{\partial Z_I} + \sum_i \frac{\partial E}{\partial c_i}
\end{align}{}
where the second sum represents the contribution of the basis set coefficients $c_i$ as they change from the reference to the target molecules. By construction\cite{def2tz,def2qz}, the first order terms for HF are zero for the def2 basis set family which might contribute to the good performance in the context of APDFT.

As discussed above, we treated the alchemical transmutation  HAl $ \leftrightarrow $ HeMg  independently. In that case, the best results were obtained  with def2-QZVPP basis set, as shown in the SI. In this case, the basis set error is 100\,mHa, one order of magnitude smaller than the error for any other basis set.

%% file: 3C_ao_correction.tex
\subsection{Correcting the Basis Set Derivative Error} \label{AOcorrection}
\begin{figure}[t]
    \includegraphics[width= \linewidth]{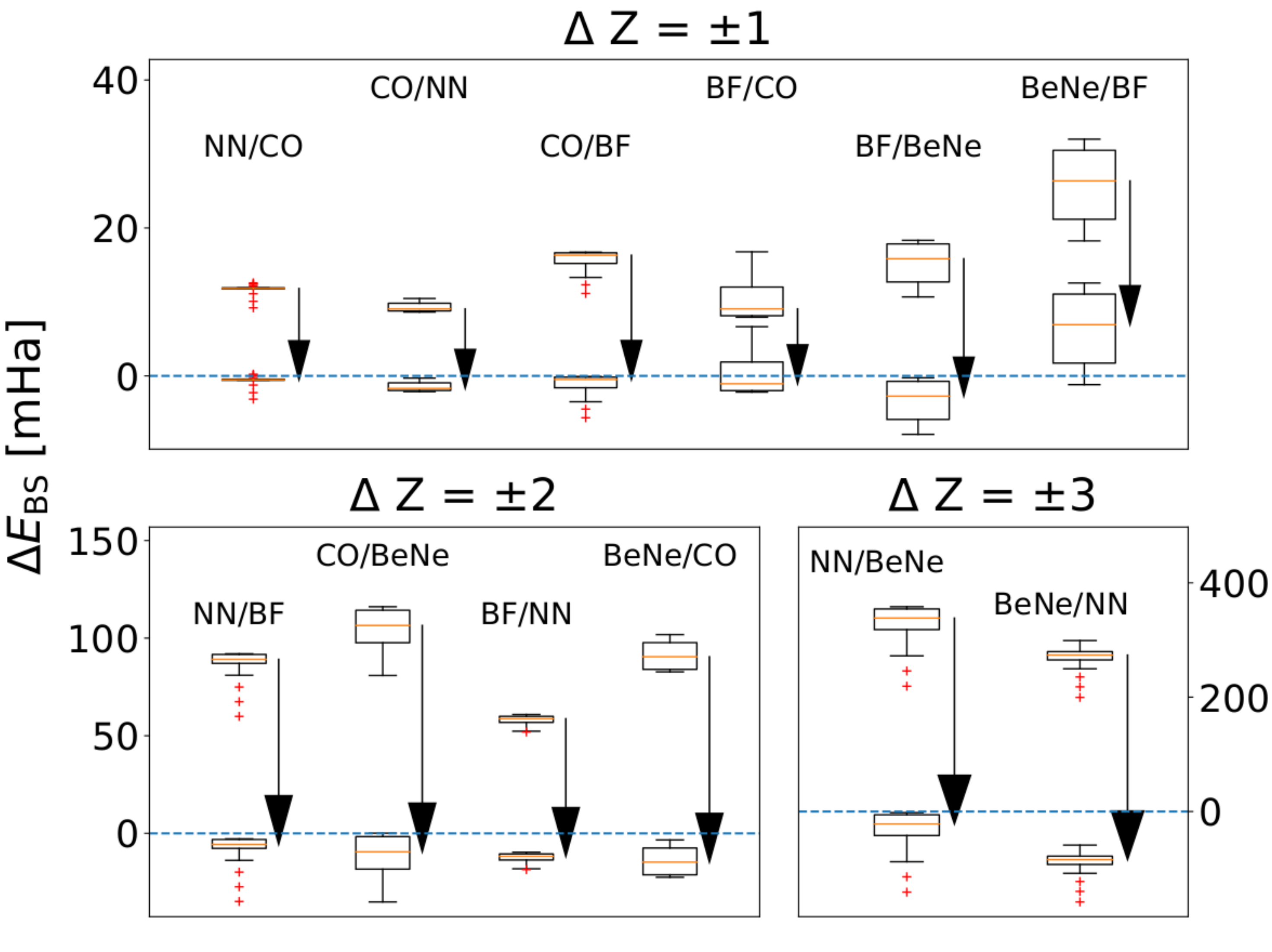}
    \caption{Systematic reduction in signed error $E^\text{T[T]}-E^\text{T[R]}$ due to the basis set correction we suggest, here shown for def2-QZVPP and the diatomic pairs of the  second period. For every alchemical transmutation (labeled reference/target in the figure) different interatomic distances lead to slightly different errors shown in boxplots. The red line is the median of the distribution, and the arrows indicate the shift due to the correction outlined in section~\ref{bsc_def_sect}.}
    \label{bs_corr}
\end{figure}

As shown above, choosing appropriate atomic orbitals basis sets can significantly reduce the overall error in APDFT. Still, even in our best case (def2-QZVPP), such error is on the order of tens of milli-Hartree, while for other smaller basis sets the error is up to two orders of magnitude larger. With the simple correction (see section \ref{bsc_def_sect}), however, this error can be substantially reduced for all basis sets, as shown in Figure~\ref{bs_corr}.

Due to the inclusion of several interatomic distances for each dimer in our data set, we always have a distribution of errors that covers the thermally accessible range. Since the correction is based on free atoms, it is independent of the interatomic distances. Consequently, the correction always moves the whole error distribution. As shown in Figure~\ref{bs_corr}, the signed error is consistently improved, since the largest error source is approximately corrected for. It is remarkable that for the cases with $\Delta Z = \pm 1$, the basis set error can be reduced to few milli-Hartree for our large span of interatomic distances.
With the exception of the case BeNe $\rightarrow $ BF, the correction generally overestimates the error. In molecules the orbitals placed on different atoms can overlap (superposition). As a consequence the basis set for each atom is more complete and thus the $\Delta E_\text{BS}$ is lower for molecular species than for isolated atoms.

%% file: 3D_Truncation_error.tex
\subsection{Truncation Error}
\begin{figure}[ht]
\includegraphics [width=\linewidth]{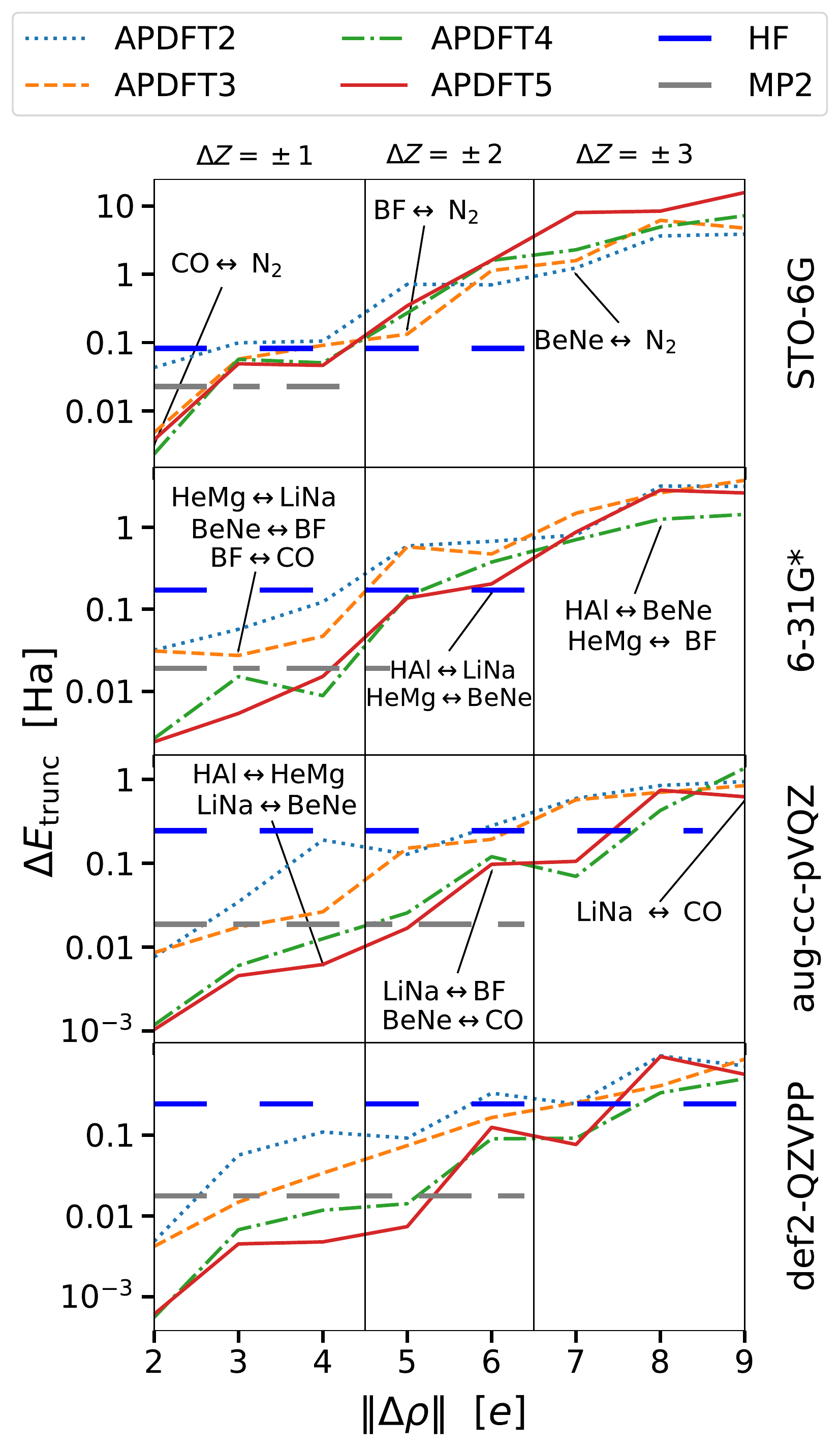}
\caption{Correlation between $ \Delta E_\text{trunc}$ and the electron density displacement $\left\lVert \Delta \rho \right\rVert$. Each panel shows the median of the error for one basis set for the different truncation orders in APDFT. For comparison, horizontal lines denote the mean absolute error of Hartree Fock and MP2 calculations, respectively. }
\label{median_error}
\end{figure} 

As with any truncated series expansion, APDFT suffers from truncation errors, i.e. the total contribution from higher order terms in the series expansion: 
\begin{align}
    \Delta E_\text{trunc} = E^\text{APDFT} -E^\text{T[R]}
\end{align}{}
An \textit{a priori} estimation of the truncation error would be highly valuable to judge the accuracy of APDFT estimations and, consequently, to define a trust region  similar to a convergence radius where the extent of the region is derived from the required accuracy with regards to the reference method.

Based on earlier work for crystal systems \cite{Chang2018}, we propose to use the integrated absolute electron density difference $\Delta \rho \equiv \rho^{\mathrm{T}}(\mathbf{r}) - \rho^{\mathrm{R}}(\mathbf{r})$ between reference and target as proxy of the error:
\begin{align}
\label{deltarho}
    \left\lVert \Delta \rho \right\rVert\equiv
    \int_\Omega d\mathbf{r} |\Delta \rho (\mathbf{r})| 
\end{align}{}

Since alchemical energy derivatives are evaluated through the alchemical perturbations of the electron density, two systems are close to each other if and only if they possess similar electron densities. In this way, a small density change in the alchemical path results in a small error. In the limit of $\left\lVert \Delta \rho \right\rVert  =0 $, reference and target are identical and the error in alchemical energy prediction is zero. Note that this integral is only weakly sensitive to changes in level of theory or basis set, as the corresponding differences in electron density are minute. This allows for simple evaluations in practice where the electron density of a low level of theory  or even  atomic densities can serve as substitute of the self-consistent electron density for the purpose of error estimation.

Figure \ref{median_error} shows that the error of APDFT correlates with $\left\lVert \Delta \rho \right\rVert$ for APDFT2 to APDFT5  for representative basis sets. The consistent trends observed in that figure allow us to give an empirical formula for the unsigned truncation error. We see that this error increases exponentially with the integrated absolute charge difference for diatomic molecules. Different expansion orders yield different steepness of this relation only.

To put the focus of the empirical error estimation on small truncation errors, we propose to fit a linear function to the logarithmized data:

\begin{equation}
\label{alpha_beta_eq}
    \log(|\Delta E_{\mathrm{trunc}}|) \approx  \alpha + \beta \left\lVert \Delta \rho \right\rVert 
\end{equation}
where the coefficients $\alpha$ and $\beta$ are found via a fit for each APDFT order and basis set. Table~\ref{alpa_beta_coeffs} shows the resulting fits. With this empirical fit, we can estimate the expected error for APDFT numbers without any additional DFT calculations. It is of great value in practise to have an expected uncertainty associated with every APDFT energy.

For more complex molecules we can consider the size-consistency of APDFT and of its error. If more than one transmutation is done on non-interacting independent sites, the total error is decomposed in a sum of site-defined contributions.

\begin{table}[ht]
    \begin{tabular*}{\linewidth}{lr @{\extracolsep{\fill}} rrr} 
    \toprule
      Basis set &  $n$ & $\alpha$ [$\log(\text{Ha})$]  & $\beta$ [$\log(\text{Ha})e^{-1}$] & $R^2$ \\ 
\hline
\midrule 
STO-6G & \rule{0pt}{12pt} 5 & -3.161 &  0.522   &  0.95     \\
6-31G* &              5 &     -3.572 &  0.480   &  0.97     \\
aug-cc-pVQZ &        5 &     -3.743 &  0.416    &  0.97     \\
def2-QZVPP &         1 &      0.047 &  0.172     &  0.88 \\
def2-QZVPP &         2 &     -2.355 &  0.273    &  0.84    \\
def2-QZVPP &         3 &     -2.806 &  0.314    &  0.98    \\
def2-QZVPP &         4 &     -3.643 &  0.388    &  0.94    \\
def2-QZVPP &         5 &     -3.973 &  0.442    &  0.92    \\
    \end{tabular*}
    \caption{The fitting parameters $\alpha$  and $\beta$ which describe the total truncation error for APDFT of order $n$ according to Eq.~\ref{alpha_beta_eq} the data are extracted from the binned values shown in Figure~\ref{median_error}. }
    \label{alpa_beta_coeffs}
\end{table}

Due to symmetry of the dimers\cite{apdft}, the error is similar for second and third order as well as for fourth and fifth order.
For small $\left\lVert \Delta \rho \right\rVert$, higher order APDFT performs substantially better than lower order ones, while for large $\left\lVert \Delta \rho \right\rVert$ the error obtained with different APDFT orders becomes comparable. We attribute this to the finite precision of the density derivatives as obtained from finite differences where for larger changes in $\Delta Z$ numerical noise is amplified.

In Figure \ref{median_error} the comparison of the truncation error between different basis sets gives results similar to those obtained for the basis set error earlier. In fact, more expanded basis sets yield lower $\Delta E_\text{trunc}$ than smaller ones and the triple and quadruple zeta basis set give very good results: for APDFT4 with $\left\lVert \Delta \rho \right\rVert < 5e $ the error is less than 10\,mHa. The best performance again was obtained with the def2 basis set: for def2-QZVPP APDFT performs better than MP2 up to $\left\lVert  \Delta \rho \right\rVert=6e$ (indicative of $\Delta Z\le 2$) and performs better than HF up to $\left\lVert \Delta \rho \right\rVert=8e$ ($\Delta Z =3$).

%% file: 3E_Vibrational.tex
\subsection{Vibrational Frequencies\label{pes_sect}}

\begin{figure}[ht]
    \centering
    \includegraphics[width=\linewidth,height=300pt]{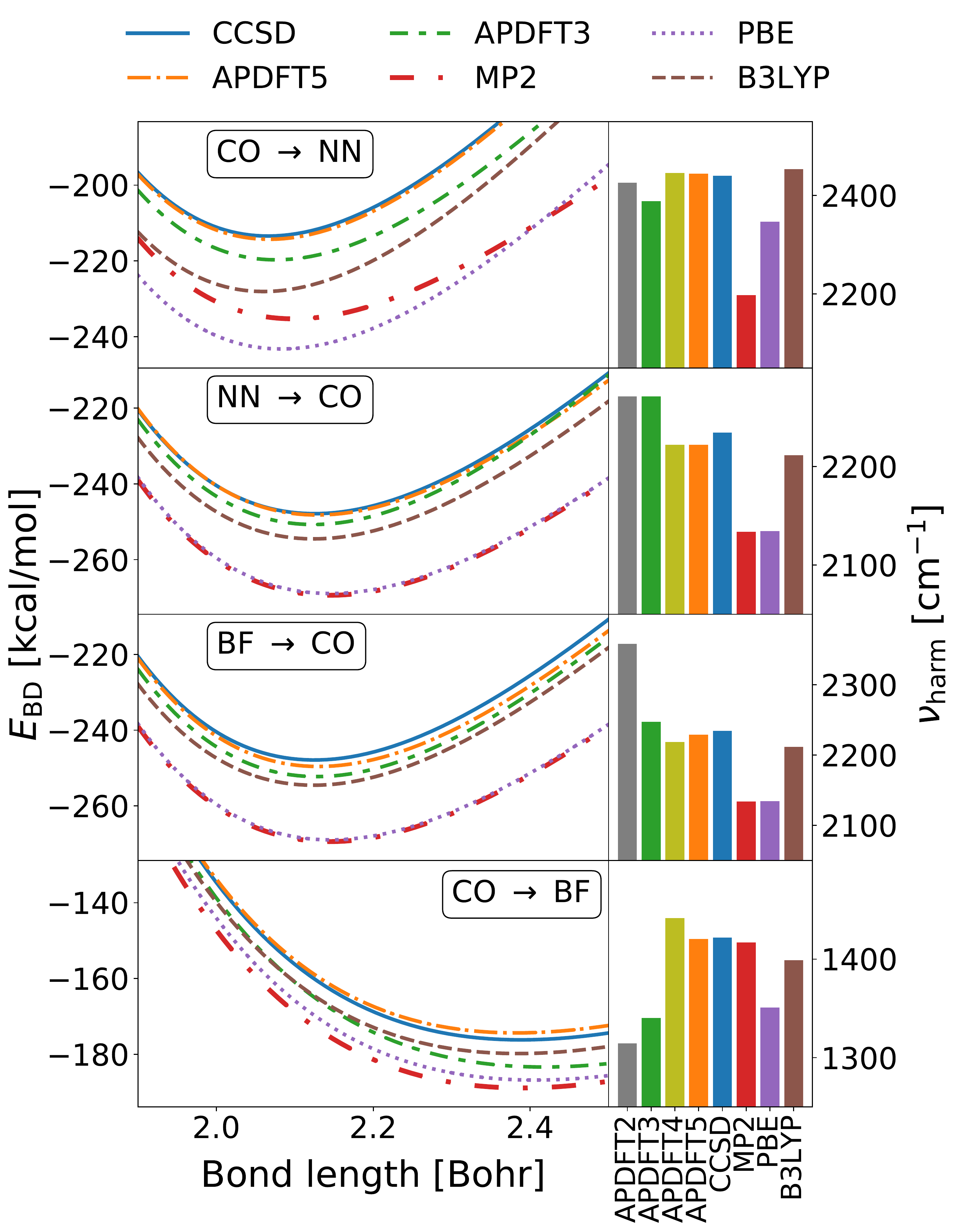}
    \caption{Homolytic bond dissociation energy $E_{\text{BD}}$ and harmonic vibrational frequencies $\nu_{\text{harm}}$ of CO, NN and BF calculated at  CCSD, MP2, HF, PBE, B3LYP and APDFT5 level of theory. For APDFT the basis set correction from section \ref{bsc_def_sect} is included. All data shown for the def2-QZVPP basis set.}
    \label{pes_curves}
\end{figure}
\begin{table}[ht]
    \begin{tabular*}{\linewidth }{lrr@{\extracolsep{\fill}}r}
\toprule
\vspace{5pt} Method & \ \ \ \ Bond Energy  & \ \ \ \ \ Bond Length  &    $\nu_\text{{harm}}$  \\ 
\midrule
\hline
\textbf{CO $\rightarrow $ NN} &\rule{0pt}{12pt} & &\\
APDFT1 &  1827.97 &  1.90 &  3118 \\
APDFT2 &  -210.76 &  2.07 &  2425 \\
APDFT3 &  -219.71 &  2.08 &  2388 \\
APDFT4 &  -214.22 &  2.07 &  2445 \\
APDFT5 &  -214.29 &  2.07 &  2444 \\
HF &  -117.32 &  2.01 &  2728 \\
MP2 &  -235.28 &  2.10 &  2197 \\
PBE &  -243.22 &  2.08 &  2347 \\ 
B3LYP &  -228.07 &  2.06 &  2453 \\  
\vspace{5pt} CCSD &  -213.43 &  2.07 &  2439 \\
\hline
\textbf{NN $\rightarrow $ CO} &\rule{0pt}{12pt} & &\\
APDFT1 &  1805.18 &  1.93 &  2954 \\
APDFT2 &  -250.73 &  2.12 &  2270 \\
APDFT3 &  -250.73 &  2.12 &  2270 \\
APDFT4 &  -248.22 &  2.13 &  2222 \\
APDFT5 &  -248.22 &  2.13 &  2222 \\
HF &  -175.58 &  2.08 &  2427 \\
MP2 &  -269.41 &  2.14 &  2133 \\
PBE &  -268.94 &  2.14 &  2134 \\
 B3LYP &  -254.52 &  2.12 &  2211 \\
 \vspace{5pt} CCSD &  -247.87 &  2.12 &  2234 \\
\hline
\textbf{BF $\rightarrow $ CO} &\rule{0pt}{12pt} & &\\
APDFT1 &  1781.15 &  1.97 &  2800 \\
APDFT2 &  -232.08 &  2.10 &  2358 \\
APDFT3 &  -252.25 &  2.13 &  2247 \\
APDFT4 &  -250.84 &  2.13 &  2218 \\
\vspace{5pt} APDFT5 &  -249.58 &  2.13 &  2228 \\
\hline
\textbf{CO $\rightarrow $ BF }& \rule{0pt}{12pt}& & \\
APDFT1 &  1893.64 &  2.08 &  2250 \\
APDFT2 &  -195.06 &  2.44 &  1314 \\
APDFT3 &  -183.34 &  2.42 &  1340 \\
APDFT4 &  -174.49 &  2.38 &  1441 \\
APDFT5 &  -174.34 &  2.38 &  1420 \\
    HF &  -137.70 &  2.35 &  1508 \\
   MP2 &  -188.87 &  2.39 &  1417 \\
   PBE &  -186.79 &  2.41 &  1350 \\
 B3LYP &  -179.79 &  2.38 &  1399 \\
   CCSD &  -176.17 &  2.39 &  1421 \\
    \end{tabular*}
    \caption{Comparison of dissociation profile in terms of homolytic bond dissociation energy [kcal/mol], bond length [Bohr] and vibrational frequencies [cm $^{-1}$] for different APDFT orders, HF, MP2, CCSD and the DFT functionals PBE and B3LYP.}
    \label{vibr_tabel}
\end{table}

Accurate estimates of potential energy surfaces  open up access to the application in vibrational spectroscopy. Even more so if the residual errors from either higher order terms or basis set intricacies are systematic around the minimum geometry configuration. \cite{Chang_bonds}
Starting from a scan of interatomic distances for one molecule, we have tested this application for APDFT energies at the same interatomic distance of a different isoelectronic molecule from our set. We then interpolated these points with a cubic spline to obtain both curvature and minimum of the target molecule. From the curvature, we calculated vibrational frequencies in the harmonic approximation.

Figure \ref{pes_curves} shows the error made by APDFT in comparison with the result that can be obtained with other computational methods. The alchemical prediction is accurate up to 2\,kcal/mol for the dissociation energies while the equilibrium bond distance is correct up to 0.01\,Bohr. Consequently, the dissociation profiles of the alchemical estimations overlay with the self-consistent CCSD profiles. In part, this is achieved thanks to the basis set correction described in section \ref{bsc_def_sect} that shifts the profile by a fixed energy, leaving minimal bond distances and the vibrational frequencies unchanged.

The dissociation profiles for other methods do not agree well with CCSD results. In particular MP2 and PBE yield errors for dissociation energies of up to 20\,kcal/mol compared to CCSD. In the right part of Figure~\ref{pes_curves} we can see that vibrational frequencies from APDFT are closer to the CCSD reference calculations than other established methods. For both energies and vibrational frequencies we observe that APDFT using perturbations of a higher level of theory can consistently outperform self-consistent results of lower levels of theory, as reported in Table~\ref{vibr_tabel}. This is particularly remarkable for vibrational frequencies which are much less affected by relative errors then total energies. Among the APDFT predictions, those are more accurate when reference and target have similar electronic structures, as example the predictions from CO to N$_2$ and \textit{vice versa} are better than the predictions from CO to BF and from BF to CO.

%% file: 3F_benzene_pyridine.tex
\subsection{APDFT derivatives on benzene \label{rings_sect} }
Finally, we show how the results from the dimer case can be transferred to larger molecules. To this end, we investigate three  targets: pyridine, pyrimidine and triazine.

With the reference molecule benzene being of D$_\text{6h}$ symmetry, and all sites being equivalent, we can reduce the number of derivatives that need to be calculated. For a APDFT3 only three derivatives of the electron density are needed to predict all targets, namely the first $\frac{\partial \rho}{\partial Z_1}$, the second $\frac{\partial^2 \rho}{\partial Z_1^2}$ and the second mixed $ \frac{ \partial^2 \rho}{\partial Z_1\partial Z_2}$, all of which were obtained by finite differences. \\

The basis set correction described in section \ref{bsc_def_sect} was obtained subtracting the CCSD energy of a nitrogen atom with its proper basis functions from the CCSD energy of an isolated nitrogen atom with the basis set of a carbon atom and a ghost hydrogen:
\begin{equation}
    \Delta E_{\text{corr}} = E^{\text{[N][C-H]}}-E^{\text{[N][N]}}
\end{equation}

The value of $\Delta E_{\text{corr}}$ was subtracted once for every CH to N transmutation in the alchemical transformation.
Based on the results in section \ref{AOerror}, we chose the def2-TZVP basis set as a compromise between cost and accuracy.
\begin{figure}[ht]
    \centering
     \includegraphics[width=\linewidth,keepaspectratio]{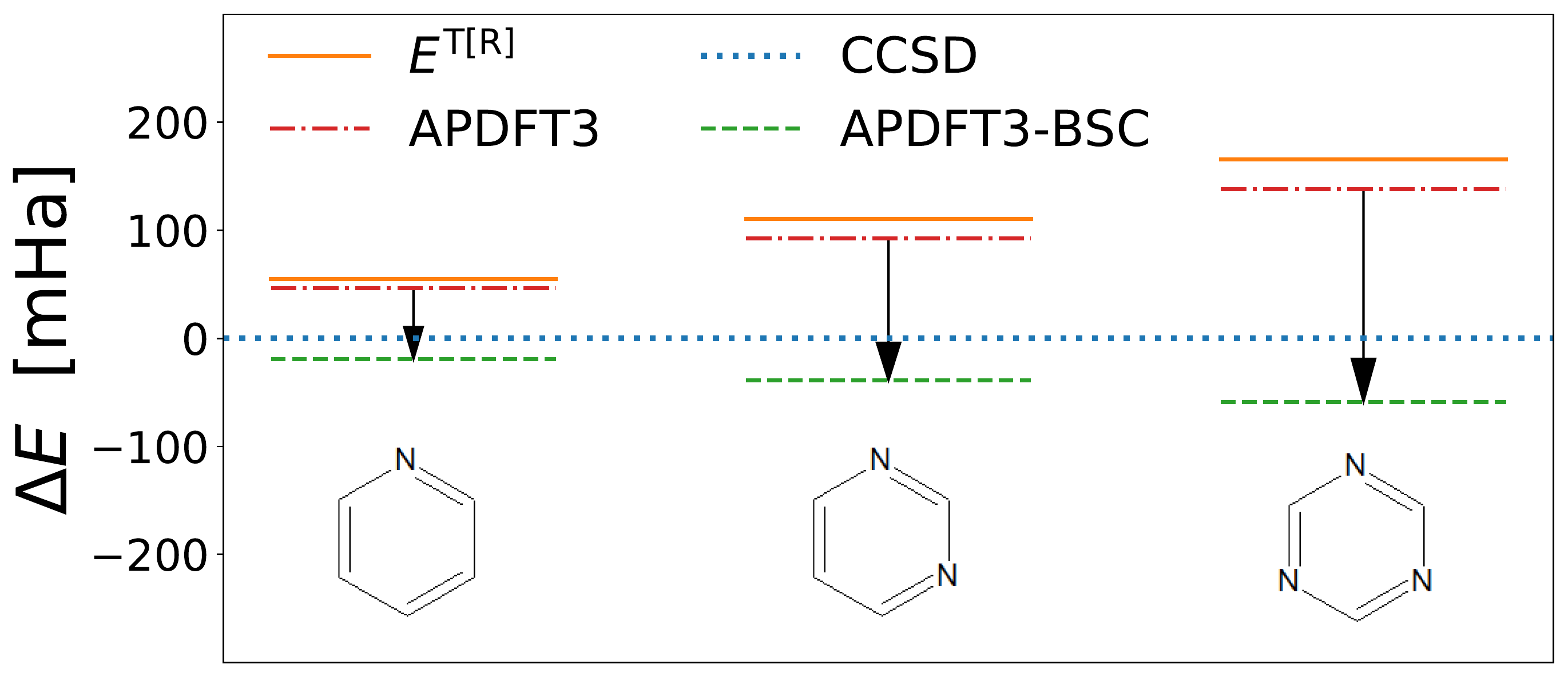}
    \caption{ $\Delta E$ in the prediction of the total energies of pyridine, pyrimidine and triazine. APDFT3 energies are calculated from a benzene reference. The results were obtained using CCSD/def2-TZVP.}
    \label{rings}
\end{figure}

\begin{table}[ht]
    \centering
    \begin{tabular*}{\linewidth}{{l @{\extracolsep{\fill}} rrr}}
    \toprule
     \vspace{5pt}& Pyridine & Bipyridine & Triazine \\
    \hline
    \midrule
     $\Delta E_\text{trunc}$ \rule{0pt}{12pt}   & -8.73 & -17.98 &  -27.74  \\
    $\Delta E_\text{BS}$   & 55.09 & 110.37 & 165.90 \\
    $\Delta E_\text{APDFT}$  & 46.35 & 92.38 & 138.16 \\
    Correction  & -65.64 & -131.28 & -196.92 \\
    Error after correction  & -19.28 & -38.90 & -58.76 \\
    \end{tabular*}
    \caption{Errors and corrections in the APDFT3 predictions of pyridine, pyrimidine, triazine from benzene, using CCSD/def2-TZVP, as shown in Figure \ref{rings}.}
    \label{tab:rings}
\end{table}

Figure \ref{rings} and Table \ref{tab:rings} show errors and corrections for the predictions of pyridine, bipyridine, and triazine. 

From the results we can see that $\Delta E_\text{BS}$ is roughly six times larger than $\Delta E_{\text{trunc}}$ and constitutes the dominating source of error, both errors increase with an approximately linear trend with the number of transmuted atoms.
The basis set correction obtained overestimates the basis set error by a sixth, as observed earlier in section \ref{AOcorrection}. The truncation error is negative in sign therefore after the correction it cancels to some degree with the error in the correction, nevertheless corrected prediction is still improved by a factor of 3.\\
If we consider the total difference in the energy of benzene and pyridine of 16.019 Ha, it is remarkable that the error with APDFT3 of 19.28\,mHa constitutes 0.12\,\% of the  total energy difference.

%% file: 4_conclusion.tex
\section{Conclusion}
In this work we analysed prediction errors in APDFT\cite{Lilienfeld2009,apdft} in terms of atomic basis set effects as well as perturbation order in the context of its application to energetics and vibrational frequencies in iso-electronic diatomics, as well as for mutating benzene to pyridine, pyrimidine, and triazine.

Our numerical results indicate that absolute energy estimates are dominated by two main sources of error: Truncation of the Taylor expansion and differences in basis sets between reference and target compound. The error due to basis set can be considerable for small basis sets, while parametrically optimized basis sets such as the Karlsruhe basis sets yield reasonably accurate alchemical derivatives. For this error source, we proposed a single atom correction which is easily implemented.

The error due to the Taylor expansion truncation has been shown to be related to the total displacement of electronic charge between reference and target molecule. We provide linear fits as an empirical relation between the two which can help to estimate the expected accuracy of an APDFT energy up to order $n = 5$ . We show that this error for small $|| \Delta \rho ||$ and for APDFT5 can  be as small as 1\,kcal/mol for the largest basis set considered. 

Regarding the prediction of vibrational frequencies, we have observed that APDFT based on CCSD calculations can reproduce the shape of dissociation curves better than GGA (at second order for $\Delta Z=1$) and B3LYP or MP2 (at fifth order). Consequently, quantities derived from these curves, such as equilibrium distances or vibrational frequencies, are also more predicted more accurately using APDFT than using DFT. 

While obtaining the alchemical derivatives from finite differences is expensive for higher orders, we showed in that good predictions can regularly be achieved already at APDFT3 level--even for vibrational frequencies. In this work, we use $n+1$ additional single points for all APDFT terms up to and including $n$-th order. APDFT becomes cost effective for problems with a large number $N $ of transmutation sites. APDFT3 requires $1+N+N^2$ single point calculations without any symmetry, while the number of possible targets increases in a combinatorial manner ($\simeq 2^N$). Symmetry can be used to further reduce the required number of reference QM calculations\cite{apdft}. This scaling behaviour renders APDFT suitable for exploring large chemical spaces from much fewer reference calculations than potential targets.